\documentclass[galaxies,article,accept,moreauthors,pdftex]{Definitions/mdpi}

\firstpage{1} 
\pubvolume{1}
\issuenum{1}
\articlenumber{0}
\pubyear{2021}
\copyrightyear{2021}
\externaleditor{Academic Editor: Wenwu Tian 
}
\datereceived{15 October 2021} 
\dateaccepted{3 November 2021} 
\datepublished{} 
\hreflink{https://doi.org/} 

\Title{Pilot study and early results of the Cosmic Filaments and Magnetism Survey with Nenufar: the Coma cluster field
}
\TitleCitation{Early Results on the Cosmic Magnetism Project with NenuFAR: the Coma Cluster field}

\Author{
	Etienne Bonnassieux $^{1,2,}${*}, 
	Evangelia Tremou $^{3,}${*},
	Julien N. Girard $^{3,}${*},
	Alan Loh $^{3,}${*},
	Valentina Vacca $^{4}$,
	Stéphane Corbel $^{5,6}$,
	Baptiste Cecconi $^{3,5}$,
	Jean-Mathias Grie\ss meier $^{5,7}$,
	L\'eon V. E. Koopmans $^{8}$,
	Michel  Tagger $^{5}$,
	\mbox{Gilles Theureau $^{5,7,9}$ and}
	Philippe Zarka $^{3,5}$
}
\AuthorNames{Etienne Bonnassieux, Evangelia Tremou, Julien N. Girard, Alan Loh, Valentina Vacca, Stéphane Corbel, Baptiste Cecconi, Jean-Mathias Grie\ss meier, L. ("Leon") V. E. Koopmans, Michel  Tagger, Gilles Theureau, Philippe Zarka}
\AuthorCitation{E. Bonnassieux, E. Tremou, J. N. Girard, A. Loh, V. Vacca, S. Corbel, B. Cecconi, J.-M. Grie\ss meier, L. V. E. Koopmans, M. Tagger, G. Theureau, P. Zarka}
\address{%
	$^{1}$ \quad Dipartimento di Fisica e Astronomia, Universita di Bologna, Via Zamboni, 33, 40126 Bologna, Italy\\
	$^{2}$ \quad INAF, Instituto di Radioastronomia, Via Piero Gobetti, 40129 Bologna, Italy\\
	$^{3}$ \quad LESIA, Observatoire de Paris, Université PSL, Sorbonne Université, Université de Paris, CNRS, 92190 Meudon, France; baptiste.cecconi@obspm.fr (B.C.); Philippe.zarka@obspm.fr (P.Z.)\\
	$^{4}$ \quad INAF{---}Osservatorio Astronomico di Cagliari, Via della Scienza 5, 09047 Selargius, Italy; valentina.vacca@inaf.it\\
	$^{5}$\quad Station de Radioastronomie de Nancay, Observatoire de Paris, PSL Research University, CNRS, Univ. Orleans, 18330 Nançay, France; corbel@discovery.saclay.cea.fr (S.C.); jean-mathias.griessmeier@cnrs-orleans.fr (J.-M.G.); tagger@cea.fr (M.T.); theureau@cnrs-orleans.fr (G.T.)\\
	$^{6}$\quad AIM, CEA, CNRS, Université de Paris, Université Paris-Saclay, 91191 Gif-sur-Yvette, France\\
	$^{7}$\quad LPC2E{---}Universite d'Orleans /  CNRS, Cedex 2, 45071 Orleans, France\\
	$^{8}$\quad Kapteyn Astronomical Institute, University of Groningen, P.O. Box 800, 9700AV Groningen, The Netherlands; l.v.e.koopmans@rug.nl\\
	$^{9}$\quad LUTh{---}Observatoire de Paris, Université PSL, CNRS, Université de Paris, 92190 Meudon, France \\
}

\corres{Correspondence: etienne.bonnassieux@unibo.it (E.B.); evangelia.tremou@obspm.fr (E.T.); julien.girard@obspm.fr (J.N.G.); alan.loh@obspm.fr (A.L.)}

\abstract{
NenuFAR, the New Extension in Nancay Upgrading LOFAR, is currently in its early science phase. It is in this context that the Cosmic Filaments and Magnetism Pilot Survey is observing sources with the array as it is still under construction---with 57 (56 core, 1 distant) out of a total planned 102 (96 core, 6 distant) mini-arrays online at the time of observation---to get a first look at the low-frequency sky with NenuFAR.
One of its targets is the Coma galaxy cluster: a well-known object, host of the prototype radio halo. It also hosts other features of scientific import, including a radio relic, along with a bridge of emission connecting it with the halo. It is thus a well-studied object.
In this paper, we show the first confirmed NenuFAR detection of the radio halo and radio relic of the Coma cluster at $34.4$ MHz, with associated intrinsic flux density estimates: we find an integrated flux value of $106.3\pm3.5$ Jy for the radio halo, and $102.0\pm7.4$ Jy for the radio relic. These are upper bound values, as they do not include point-source subtraction.
We also give an explanation of the technical difficulties encountered in reducing the data, along with steps taken to resolve them. This will be helpful for other scientific projects which will aim to make use of standalone NenuFAR imaging observations in the future.
}


\begin{document}




\def\weight#1{\omega_{#1}}
\def\weightpr#1{\omega'_{#1}}
\def\vecweight#1{\bm{\omega}_{#1}}
\def\matJ#1{\bm{{J}}_{#1}^{t\nu}}
\def\lvec{\bm{l}}
\def\uvec#1{\bm{u}_{#1}}
\def\matK#1{\bm{{K}}_{#1,\lvec}^{t\nu}}
\def\matAa#1{{{A}}_{#1}^{t\nu}}
\def\matA#1{{{A}}_{#1,\lvec}^{t\nu}}
\def\matAo#1{{{A}}_{#1,0}^{t\nu}}
\def\matAlo#1{{{A}}_{#1,\lvec_0}^{t\nu}}
\def\k#1{k_{#1,\lvec}^{t\nu}}
\def\ko#1{k_{#1,0}^{t\nu}}
\def\klo#1{k_{#1,\lvec_0}^{t\nu}}
\def\matV#1{\bm{V}_{#1}^{t\nu}}
\def\matVhat#1{\hat{\bm{V}}_{#1}^{t\nu}}
\def\matB{\bm{B}}
\def\matBl{{\matB}_{\lvec}^{\nu}}
\def\matBlo{{\matB}_{\lvec_0}^{\nu}}
\def\I{\mathbf{I}}
\def\dirtyIpq{I_{\text{dirty}}^{pq,t\nu}}
\def\dirtyIpqsmr{\tilde{I}_{\text{dirty}}^{pq,t\nu}}
\def\dirtyI{I_{\text{dirty}}}

\section{Introduction}\label{sec.intro}


The Coma galaxy cluster is not only one of the first historically observed galaxy clusters \citep{1951PASP...63...61Z}, but also the first galaxy cluster with the detection of either a radio halo \citep{1959Natur.183.1663L} or radio relic \citep{1981A&A...100..323B,1991A&A...252..528G} in the literature. It is very well-studied at a range of frequencies \citep{2021ApJ...907...32B}. As an object known to host several physical components of considerable scientific interest, it was chosen as the first galaxy cluster to be observed as part of the NenuFAR Cosmic Filaments \& Magnetism Pilot Survey project.

NenuFAR, the New Extension in Nançay Upgrading LoFAR \citep{nenufar},  is a very powerful instrument optimised to offer unparalleled sensitivity in the Northern sky at the lowest astronomically-relevant frequencies (10--85 MHz) which can be probed from the Earth.
In its imager mode, NenuFAR features a dense network of short baselines (down to 25 m), making it exceptionally powerful for detecting diffuse emission on large scales. However, its longest baselines are still quite short (3 km for the ``distant'' mini-arrays once complete); this means that it offers rather poor angular resolution ($20'$ at the time our observation was taken, with 56 of a total planned 96 core mini-arrays online and only 1 of a planned 6 distant mini-arrays operational). This means that its sensitivity, in its standalone imager mode, is strongly limited by the confusion noise (i.e. the ``noise'' introduced by the presence of ``clouds'' of faint sources which cannot be resolved by the instrument).

In this paper, we present and comment the first confirmed detection of the Coma cluster made using NenuFAR at decametric wavelengths. 
In Section~\ref{sec.1}, we discuss the data reduction process for our field, the difficulties associated with wide-field, low-resolution imaging with NenuFAR, and the self-calibration behaviour of our data.  In Section~\ref{sec.2}, we discuss the final image itself. Finally, in Section~\ref{sec.conclusion} we close out with a discussion and future prospects, both for this field and the pilot survey more broadly.

\section{Observations, Data Reduction, and Validation}\label{sec.1}
\unskip

\subsection{Observations and Data~Reduction}\label{sec.1.1}

Our data were observed in the framework of the Cluster Filament \& Cosmic Magnetism Pilot Project Key Programme (ES09). on 16/12/2020, as part of a 2-hour observation centred on Virgo A between 07:00 and 09:00 UTC. Their associated NenuFAR observation ID (OBS\_ID) is 20201216\_070000\_20201216\_090300\_COMA\_PILOT00. The data were initially recorded at 3 kHz frequency resolution and 1s time resolution, with a bandwidth ranging from 29.80 MHz to 70.03 MHz. These data were subsequently flagged for radio-frequency interference (RFI) using \texttt{AOFlagger} \citep{2010ascl.soft10017O}, and finally averaged to $25$ kHZ frequency resolution and $1$ second time resolution as part of the standard NenuFAR pre-processing approach. 
Despite the wide bandwidth of this observation, we currently present results using the band ranging from $29.80$ MHz to $38.98$ MHz. The analysis of the  larger-bandwidth data will be the scope of future work.


The initial calibration was done using a sky model which included only Virgo A, modelled to first order as an unresolved point source located at Virgo A's coordinates and with the integrated flux value given by \citet{2017ApJS..230....7P}. The calibration was done using \texttt{DPPP} \citep{2018ascl.soft04003V}, with a time interval of $8$ s and a frequency interval of $200$ kHz (corresponding to intervals of 8 measurements in time and 8 measurements in frequency per calibration solution - we therefore do not track frequency changes within a given sub-band). The Jones matrix solved for was a diagonal matrix, and the calibration solutions were applied in both amplitude and phase. The imaging was done using \texttt{WSClean}'s multi-scale cleaning \citep{2014MNRAS.444..606O} with a cell size of $4\mathrm{'}$, which corresponds to a point-spread function (PSF) oversampling factor of 5; the restoring beam size is consequently 20$\mathrm{'}$. A Briggs weight with robust value of $-0.5$ was used as a compromise between the constraints imposed by noise statistics and PSF conditioning.

After this, a sky model was extracted from the resulting image using \texttt{PyBDSF} \citep{2015ascl.soft02007M} and the process was repeated using it as the initial sky model for calibration - a process otherwise known as self-calibration. This was done iteratively until the differences in the images between two passes of self-calibration became negligible: in our case, this meant 10 iterations. Detailed tests and comparisons are given in Section~\ref{sec.1.2}. The final image is shown and discussed in Section~\ref{sec.2}.

\subsection{Technical Tests and Validation}\label{sec.1.2}

\subsubsection{Calibration}

Although the basic calibration parameters were described above, multiple software suites were tested to verify our results. Notably, calibration (with equivalent parameters) was performed both by \texttt{DPPP} \citep{2018ascl.soft04003V} and \texttt{killMS}, the calibrator package of the Wirtinger \mbox{pack \citep{2015MNRAS.449.2668S,2018A&A...611A..87T}}.

We found that, while both suites gave completely equivalent gain solutions, \texttt{DPPP} was significantly faster than \texttt{killMS} on the Centre de Données de Nancay (CDN) computational cluster. We did not perform direction-dependent calibration. We expect that this would improve the final image quality: ionospheric effects (which are direction-dependent effects) are very strong at the low observing frequencies of NenuFAR. At the time of writing, however, the lack of an integrated NenuFAR beam model in imaging and calibration suites provides an additional source of direction-dependent errors, and we aim to implement such an integrated beam model before attempting direction-dependent calibration.

\subsubsection{Imaging}

Because the NenuFAR field of view is so large (nominally $26^\circ$ at 30 MHz), modern imaging suites are required. We tested both \texttt{WSClean} \citep{2014MNRAS.444..606O} and \texttt{DDFacet} \citep{2018A&A...611A..87T}, both able to handle creating images with a FoV of $45^\circ$, which is necessary considering that imagers often make use of a padding parameter to avoid edge effects in interferometric images.

\begin{figure}[H]
	\includegraphics[width=0.9\linewidth]{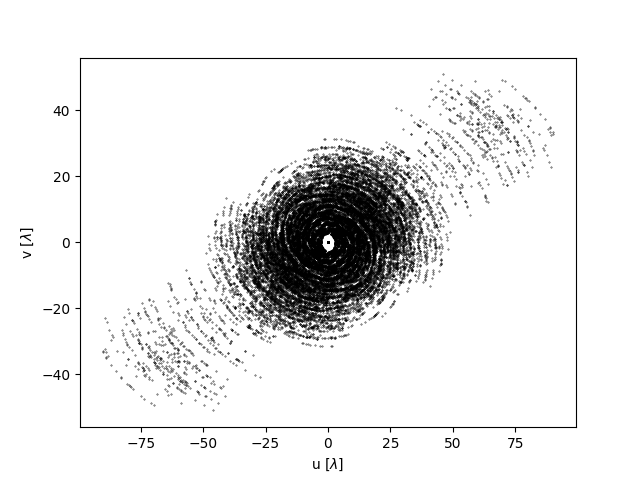}
	\caption{Achromatic $uv$-coverage of our NenuFAR observation, in units of central wavelength (in this case, the associated central frequency is 29.8 MHz). These are then modulated by our observing frequencies at each measurement. Note that the axes only extend to 75$\lambda$.} \label{fig.uvcoverage}
\end{figure}

\begin{figure}[H]
	\includegraphics[width=0.9\linewidth]{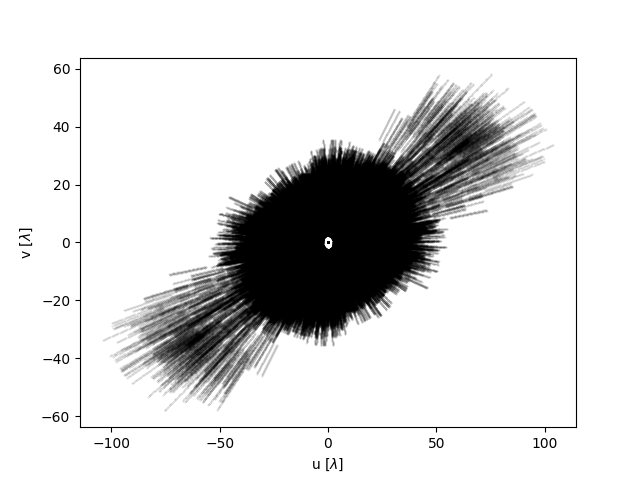}
	\caption{Chromatic $uv$-coverage of our NenuFAR observation, in units of central wavelength. The relevant frequencies are 29.8 MHz to 38.8 MHz. Note that the axes now extend to 100$\lambda$. Note the extremely high point density in this plot: the high fractional bandwidth ensures a very complete coverage for all non-distant miniarrays, and even fill out the $uv$-gap between the core and distant baselines.} \label{fig.uvcoverage.freq} 
\end{figure}

Although both suites seemed to give equivalent results, \texttt{WSClean} was found to converge faster overall. Neither imager managed to converge towards particularly deep images, which we believe is partly due to the poor overall $uv$-coverage of our data, shown in Figures~\ref{fig.uvcoverage} and \ref{fig.uvcoverage.freq}: 2 h observation leading to the strongly elliptical restoring beam which can be seen in the images in Section~\ref{sec.2}. The lack of distant mini-arrays also leads to a high confusion noise ($\sim 2$ Jy), which we believe limits the current observations.

NenuFAR mini-arrays each produce a broad beam of $(700/\nu)^\circ$ with $\nu$ in MHz. The full core array provides an angular resolution of $(43/\nu)^\circ$, which can improve to $\sim(6/\nu)^\circ$ with the distant mini-arrays, with $\nu$ again in MHz. All these values are taken from \citet{nenufar}. Its $8^\circ-46^\circ$ FoV ($\sim26^\circ$ at 30 MHz) may however also cause some of the approximations usually made in interferometric theory to fail notably the small-field approximation, cf Section 3 of  \citep{2011A&A...527A.107S} which could fail as early as 8 degrees away from phase centre. If that is the case, then exploiting NenuFAR data to its fullest will require significant theoretical developments. At the time of writing, we find that existing tools appear able to handle this issue at 30 MHz.

\subsubsection{Model~Updating}

Performing self-calibration requires updating the sky model for the field of interest in order to provide a better starting point for subsequent calibration attempts. Compromises must be found between reducing computing time by e.g., simply writing the deconvolution model to disk during imaging, and minimising sources of error by e.g., removing calibration/imaging artefacts from the model before calibration.

Generally, if the image can be treated as a reliable model for the sky brightness distribution being observed, it is best to simply use the predicted visibilities associated with our best imaging model. If it cannot, it is best to use packages such as \texttt{PyBDSF} to create a conservative sky model.

We have tested both approaches and found that they tend to converge towards a similar sky brightness distribution in general with sufficient iterations of self-calibration. However, we also find that, in general, diffuse emission in the field is only partially picked up during imaging. As such, we recommend using the conservative, \texttt{pyBDSF}-based approach if the field includes significant diffuse emission, as this will begin by minimising the calibration and imaging artefacts around the bright point sources and therefore lead to the diffuse emission being added more faithfully into the model once the signal-to-noise ratio for it is sufficient, rather than being added iteratively. Although both converge in our case, this conservative approach helps to diagnose what is happening should the self-calibration process diverge. In our case, the default \texttt{PyBDSF} threshold values were used: $5\sigma$ for pixels to be taken into account, and $3\sigma$ for islands to be created.

Finally, we note that we have started here with a simple model of Virgo A, and performed our entire self-calibration process in terms of apparent flux---never taking the NenuFAR response into account in the process. The conversion of apparent to intrinsic flux is thus the next point of discussion.

\subsubsection{Flux~Scaling}

Because interferometers do not measure the zero-scale Fourier term (which corresponds to antenna autocorrelations, typically discarded due to observational effects such as radio frequency interference), a flux scale calibrator is typically needed. We use Virgo A, which lies within our FOV, for this purpose, extrapolating its flux density from \citet{2017ApJS..230....7P}. Although \citet{2012MNRAS.423L..30S} do extend their flux scale below 50 MHz, they do not give explicit parameter values for Virgo A (3C273). Because \citet{2017ApJS..230....7P} is designed to be compatible with this flux scale at low frequencies, we proceed with its integrated flux model for this paper. This results in a $5\%$ flux scale error.
In addition, in order to recover physical flux density measurements of sources away from the pointing centre, the image must be corrected by the instrument's response at that point in the sky. 
At the time of writing, the NenuFAR beam model is not implemented in either of the two tested imaging or calibration suites. However, there exists a standalone implementation which modulates source catalogues with the NenuFAR beam given a specific dataset from which to read relevant observational parameters such as central frequency, bandwidth, pointing center, and positions of target sources. It is described in \citet{alan_loh_2020_4279405}.

We used this implementation to calculate the beam response at Virgo A's position for our observation. Because this source is a calibrator listed in \citet{2017ApJS..230....7P}, we then only had to extrapolate its known intrinsic flux down to $30$ MHz. Although this observing frequency lies outside the domain of applicability of the cited flux scale, the spectral behaviour of Virgo A's integrated flux is very stable as a function of frequency. We therefore feel that this approach is valid.

The intrinsic flux density of Virgo A is thus taken to be to $3720\pm186$ Jy, following the $5\%$ flux scale error stated in \citet{2017ApJS..230....7P}. This value is then modulated by the value of the NenuFAR beam response at the position of Virgo A to then determine the flux scale used for the Coma cluster itself.

\section{Image and Analysis}\label{sec.2}

In this section, we present the images made using the NenuFAR data described thus far. We begin with an overview of the whole field of view, and then focus on specific sources of interest in the field. In all cases, our flux measurements are performed using \texttt{CASA} version 5.6.0-60 \citep{2007ASPC..376..127M} 
  using the task \texttt{imfit}.

\subsection{Overall~Field}\label{sec.2.0}

\begin{figure}[H]
	\includegraphics[width=\linewidth]{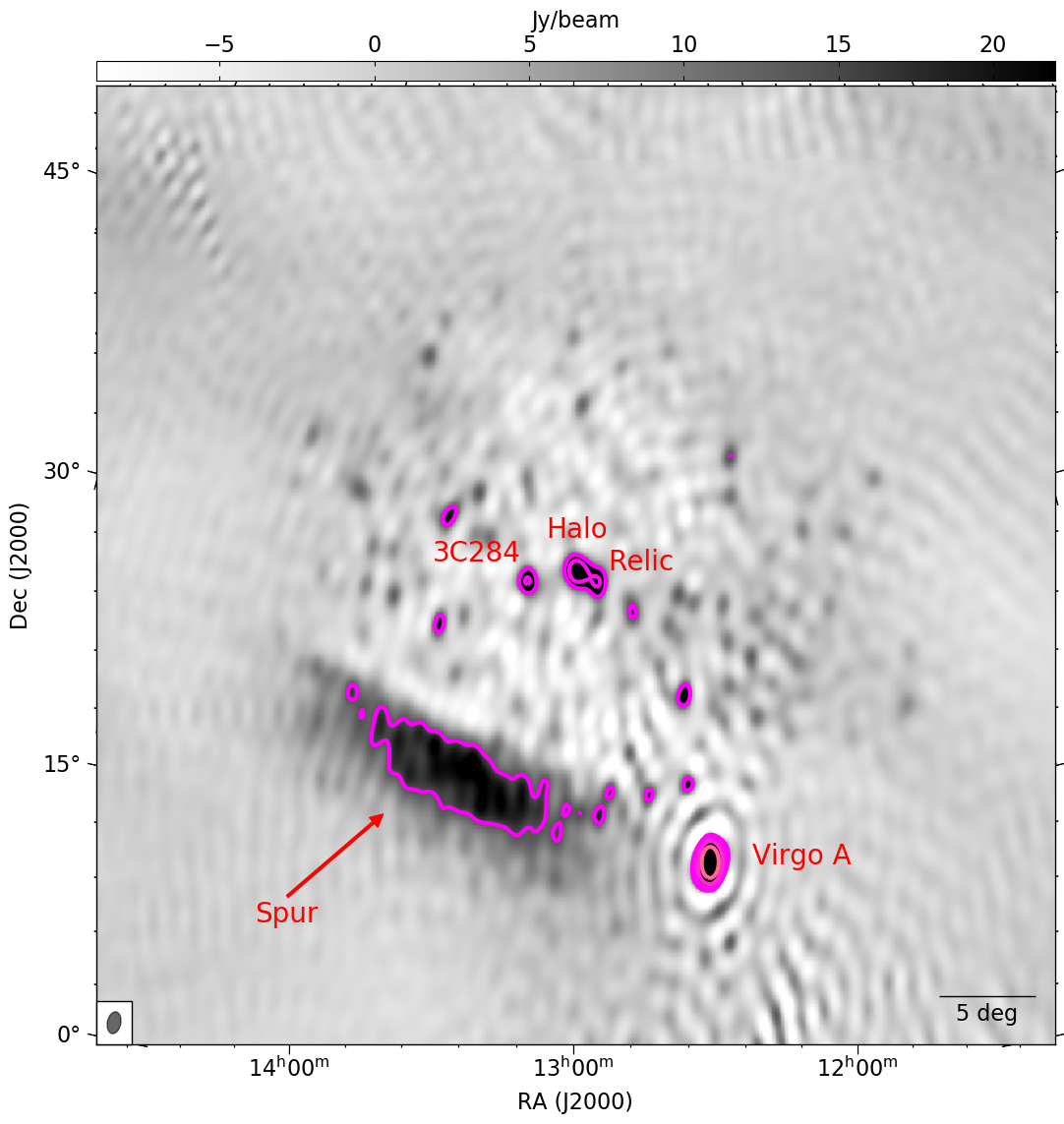} 
	\caption{Apparent sky brightness distribution image of the NenuFAR field of view, centred on the Coma cluster, 2 hour integration time. Green corresponds to negative flux values, i.e. artefacts. The noise near Coma is $2$ Jy beam$^{-1}$, and the pink contours are the image overlaid on itself starting at $3\sigma$.. The restoring beam size is shown in the bottom left.} \label{fig.widefield}
\end{figure}

The full field of view of our observation is given in Figure~\ref{fig.widefield}. Note that this is an apparent flux image, where the sources are modulated by the instrument's response as a function of their distance from pointing centre. The Galactic coordinates for the Coma cluster, which lies at the center of this field, are ($l=$58:04:44.79, $b=$87:57:27.81). All flux values are calculated by measuring the integrated flux density within regions, with the associated RMS, and converting from Jy beam$^{-1}$ to Jy.

We see that Virgo A (12$^h$30$^m$49$^s$, +12$^\circ$ 23' 28''), the Coma cluster (12$^h$59$^m$48.7$^s$, 	+27$^\circ$ 58' 50''), a scattering of sources, and a very bright and large-scale diffuse emission ($10^\circ\times5^\circ$) associated with the North Galactic radio spur are present in our field. The latter is detected both at the $5\sigma$ and the $3\sigma$ levels, though our overlays show the extent of the source at $3\sigma$ significance. This latter source will not be discussed further in this paper, save to note that it is undoubtedly present in the data itself, not the result of bias in calibration or imaging, and that it will be the subject of a future paper dedicated to its multi-$\lambda$ analysis.

\subsection{Virgo~A}\label{sec.2.1}

Lying about 17 degrees away from the Coma cluster, Virgo A is nevertheless still within the NenuFAR field of view. We therefore use it both as a starting point for our data reduction process, and the validator for our final flux scale. We find no significant astrometric offset between the position of Virgo A in our image and that given in the literature (e.g., \citep{2017ApJS..230....7P}), and we measure a dynamic range near Virgo A of only about 34---this is consistent with the clear presence of artefacts around this source.
\subsection{The Coma~Cluster}\label{sec.2.2}


\begin{figure}[H]
	\includegraphics[width=\linewidth]{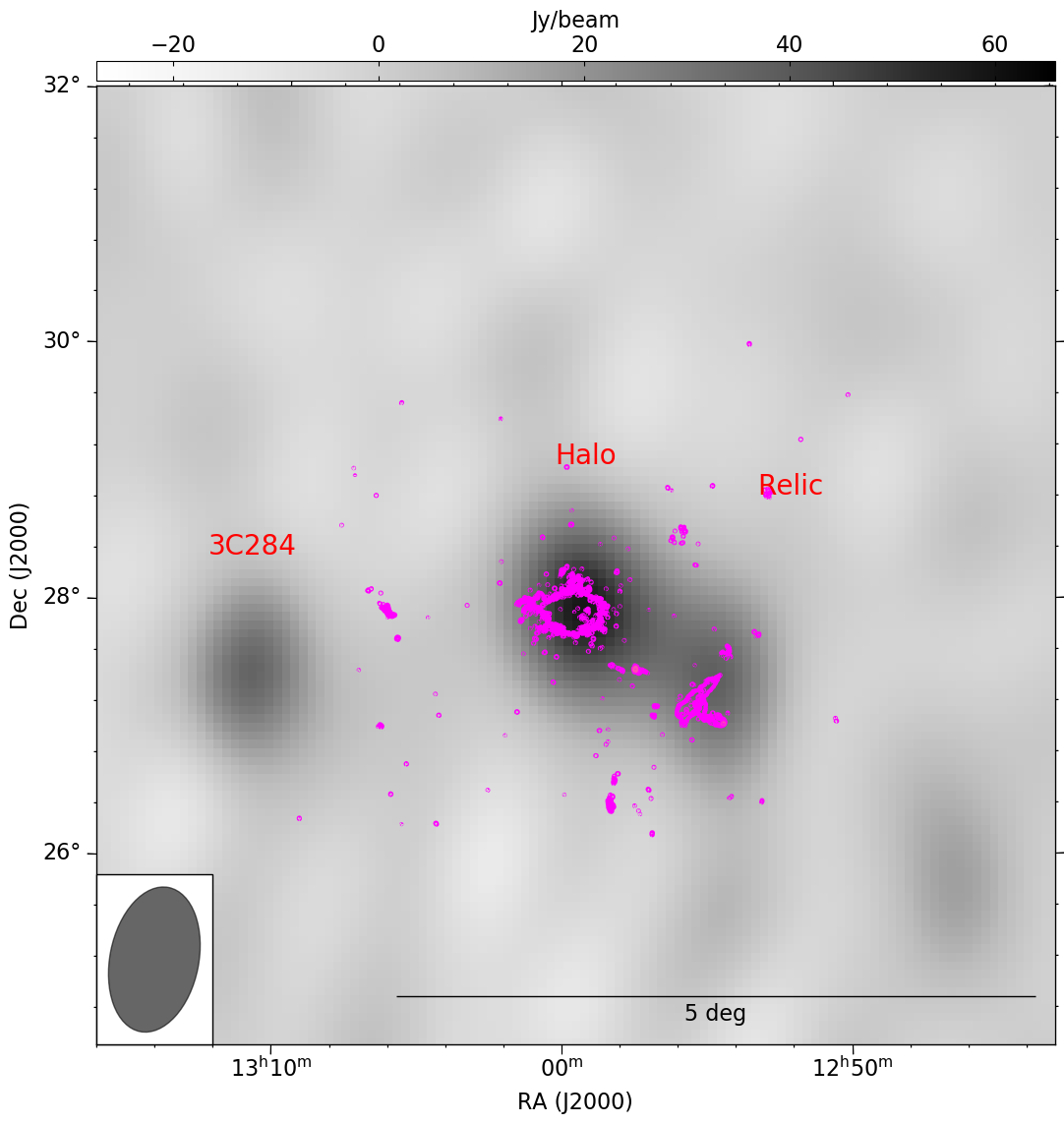}
	\caption{Intrinsic flux image of the Coma cluster. Green corresponds to negative flux values, i.e. artefacts and confusion noise. The noise is $3.5$ Jy beam$^{-1}$, and the pink contours correspond to an overlay of the 144 MHz LOFAR image of Coma shown in Fig. 1 of \citet{2021ApJ...907...32B}, starting at $5\sigma$. The integrated flux density of the Coma cluster itself (halo and point sources) is $81.5\pm15$ Jy. The restoring beam size is shown in the bottom left.} \label{fig.coma} 
\end{figure}

We see in Fig. \ref{fig.coma} that the source at the centre of our NenuFAR image does indeed correspond to the known structure of the Coma cluster as seen with LOFAR. The source to the West of the cluster is 3C284, a bright radio galaxy which was subtracted from the data imaged by \citet{2021ApJ...907...32B}. We measure an integrated flux density of $66.0\pm4.1$ Jy for this source. All three of these component are therefore known objects and can be considered successful NenuFAR detections.

As for the flux measurements, we find that the Coma cluster core (radio halo and point sources) has an integrated flux density of $106.3\pm3.5$ Jy while the radio relic to its South-West (including embedded point sources) has a total flux density of $102.0\pm7.4$ Jy. Because our application of the NenuFAR beam response over our observation's bandwidth is only accurate to first order, these can only be treated as first-order measurements. They are consistent with the literature if we take into account the fact that point source subtraction was not performed in our case: \citet{1989AJ.....97.1561H} reports a flux density measurement of $41 \pm 10$ Jy for the Coma cluster at 30MHz, although this value estimates and subtracts the contribution from embedded point sources. It is therefore unsurprising that our own flux measurement is higher, as we have not done so. We note that this measurement is shown in Fig. 5 of \citet{2003A&A...397...53T} where it clearly lies below the power-law line followed by the remainder of the data points below 1 GHz. It is likely, then, that our measurement provides an upper bound of the flux density of the Coma radio halo at 30 MHz, where \citet{1989AJ.....97.1561H} provides a lower bound using the Clark Lake radio telescope. As we have not performed point source subtraction within the halo, we cannot comment further.

\section{Conclusions and Future~Work}\label{sec.conclusion}

We have reduced a 2-h observation of the Coma cluster using NenuFAR in its imager mode, and find that it measures an integrated flux density of 81.5 ± 15 Jy for the radio halo,and 38.1 ± 8 Jy for the radio relic. Both components are clearly identified and present in our NenuFAR image above a $5\sigma$ significance threshold.

Multiple tests were run over the course of reducing these data, with the conclusion that NenuFAR data reduction in this regime is currently very strongly constrained by the data itself. This means that choices in calibration and imaging parameters, sky model generation, etc have little impact on the final sky brightness distribution, though they can significantly affect the number of iterations required to converge to a final result: consequently, we recommend using at least 10 iterations of self-calibration when reducing NenuFAR data.


Other sources are present in the field and detected at $5\sigma$ significance level, including large-scale diffuse emission from the North Galactic spur. The full scientific analysis of the field will be the topic of a future paper. On the technical side, implementing the NenuFAR beam directly into imaging and calibration software will help reduce the intrinsic flux measurement errors, as well as improving the conditioning of the calibration and imaging overall. We aim to do this as more mini-arrays continue to come online.



Overall, this is a very promising start. NenuFAR can operate as a standalone instrument to deliver reasonable measurements of sources of interest despite its low resolution. Longer observations (to improve the $uv$-coverage as more distant mini-arrays, 3 km away from the core, are built), combined with the dual use of NenuFAR as a standalone instrument and as a LOFAR superstation, should go a long way towards opening up an entirely new set of possibilities: with high sensitivity on both long and short baselines, the source subtraction process will become much easier, and the possibility of a high-resolution, large-FoV survey instrument at 30 MHz is becoming ever closer to reality. More NenuFAR mini-arrays are being built at the time of writing, and as soon as 3 distant mini-arrays come online, we aim to perform full 8-hour observations of the 5 target fields of the Galaxy Clusters, Filaments and Cosmic Magnetism Pilot Survey.

Even at its current stage, with 57 (56 core, 1 distant) out of a total planned 102 (96 core, 6 distant) built, we have observations which reliably converge towards a sky brightness distribution with known features, and it does so with a minimal amount of user involvement - the self-calibration process used in this work is fully automated. As such, we expect that we will be able to quickly and easily determine whether we can in fact detect emission from the cosmic filaments at 30 MHz, once the distant mini-arrays are built and we meet the confusion noise threshold required by our science case to detect filamentary emission. The primary aims of our Pilot Survey are thus well on track to be fulfilled.

\vspace{6pt} 

\authorcontributions{
E.B.; Data Curation: B.C., A.L., E.T.; Formal Analysis:  E.B., E.T.; Funding Acquisition: S.C., B.C., J.-M.G., M.T., G.T., P.Z.; Methodology: E.B.; Project administration: E.B.; Resources:  B.C., A.L., E.T., J.N.G., S.C., J.-M.G., M.T., G.T., P.Z.; Software: A.L., E.B.; Validation: E.B., A.L., E.T., J.N.G.; Visualization: E.B., E.T.; Writing---original draft: E.B.; Writing---review \& editing: E.B., E.T., J.N.G., A.L., V.V., S.C., B.C., J.-M.G., L.V.E.K., M.T., G.T., P.Z. All authors have read and agreed to the published version of the manuscript.
}

\funding{E. Bonnassieux acknowledges support from the ERC-Stg grant DRANOEL, No. 714245.  This paper is based on data obtained using the NenuFAR
	radio-telescope. The development of NenuFAR has been supported by personnel
	and funding from: Station de Radioastronomie de Nançay, CNRS-INSU,
	Observatoire de Paris-PSL, Université d’Orléans, Observatoire des Sciences
	de l’Univers en Région Centre, Région Centre-Val de Loire, DIM-ACAV and
	DIM-ACAV+ of Région Île-de-France, Agence Nationale de la Recherche. We
	acknowledge the use of the Nançay Data Center computing facility (CDN – Centre
	de Données de Nançay). The CDN is hosted by the Station de Radioastronomie
	de Nançay in partnership with Observatoire de Paris, Université
	d’Orléans, OSUC and the CNRS. The CDN is supported by the Région Centre Val de Loire, département du Cher. The Nançay
	Radio Observatory is operated by the Paris Observatory, associated with the
	French Centre National de la Recherche Scientifique (CNRS). VV acknowledges support from INAF mainstream project “Galaxy Clusters Science with LOFAR” 1.05.01.86.05.1}

\institutionalreview{Not applicable.}

\informedconsent{Not applicable.}

\dataavailability{The datasets used for this work will be made publicly available from the Nenufar-DC once it comes online, and can then be found using the OBS\_ID cited in the paper. Until that point, it may be accessed by contacting CDN (\url{contact_cdn@obs-nancay.fr},  
Data-products such as the final science images will also be accessible in this way.}

\conflictsofinterest{The authors declare no conflict of interest.} 



\end{paracol}


\printendnotes[custom]

\reftitle{References}


\end{document}